%% file: cogsci25.tex
\documentclass[10pt,letterpaper]{article}

\usepackage{cogsci}

\usepackage{times}
\usepackage{soul}
\usepackage{url}
\usepackage[hidelinks]{hyperref}
\usepackage[utf8]{inputenc}
\usepackage{graphicx}
\usepackage{amsmath}
\usepackage{amsthm}
\usepackage{booktabs}
\usepackage{algorithm}
\usepackage{algorithmic}
\usepackage[switch]{lineno}
\usepackage{amsfonts}
\usepackage{multirow}

\cogscifinalcopy % Uncomment this line for the final submission 

\usepackage{pslatex}
\usepackage{apacite}
\usepackage{float} % Roger Levy added this and changed figure/table
\title{Towards a Vision-Language Episodic Memory Framework:\\ Large-scale Pretrained Model-Augmented Hippocampal Attractor Dynamics}

\author{
	\begin{tabular}{cccc}
		\begin{tabular}{c}
			{\large \bf Chong Li} \\
			{\normalsize lichong23@m.fudan.edu.cn}\\
			Fudan University
		\end{tabular}
		&
		\begin{tabular}{c}
			{\large \bf Taiping Zeng\thanks{Corresponding author.}} \\
			{\normalsize zengtaiping@fudan.edu.cn}\\
			Fudan University
		\end{tabular}
		&
		\begin{tabular}{c}
			{\large \bf Xiangyang Xue} \\
			{\normalsize xyxue@fudan.edu.cn}\\
			Fudan University
		\end{tabular}
		&
		\begin{tabular}{c}
			{\large \bf Jianfeng Feng} \\
			{\normalsize jffeng@fudan.edu.cn}\\
			Fudan University
		\end{tabular}
	\end{tabular}
}

\begin{document}
	
	\maketitle
	\begin{abstract}
Modeling episodic memory (EM) remains a significant challenge in both neuroscience and AI, with existing models either lacking interpretability or struggling with practical applications. This paper proposes the Vision-Language Episodic Memory (\textbf{\textit{VLEM}}) framework to address these challenges by integrating large-scale pretrained models with hippocampal attractor dynamics. VLEM leverages the strong semantic understanding of pretrained models to transform sensory input into semantic embeddings as the neocortex, while the hippocampus supports stable memory storage and retrieval through attractor dynamics. In addition, VLEM incorporates prefrontal working memory and the entorhinal gateway, allowing interaction between the neocortex and the hippocampus. To facilitate real-world applications, we introduce EpiGibson, a 3D simulation platform for generating episodic memory data. Experimental results demonstrate the VLEM framework’s ability to efficiently learn high-level temporal representations from sensory input, showcasing its robustness, interpretability, and applicability in real-world scenarios.
		
	\end{abstract}
	
	\section{1.\ Introduction}
	
	% background: the importance of episodic memory for understanding the human brain and intelligence, and applications
	% episodic memory progress, what is important
	% current models: classic models, such as Hopfield, and deep learning models, point out disadvantages
	% we propose
	% contribution
	
	%	1.episodic-memory 对人工智能和脑科学有很重要的意义，在脑科学里有很大的进展，很多人尝试去模仿episodic-memory
	%   2.这些工作有什么问题--->我们要做什么工作

	% EM对通用人工智能的重要性 3~4行 从AI引入EM
	%AI好->AI有...问题--->EM的意义
	
	The rapid progress in AI has led to the language models that can produce texts almost indistinguishable from human writing~\cite{GPT-semantic}, representing a major step forward in realizing semantic memory functions~\cite{Kumar2020SemanticMA}. However, modeling episodic memory\textemdash another key type of memory related to personal experiences\textemdash remains a significant challenge.
	
	Episodic memory (EM) refers to the ability to store and consciously recall specific memories of past events~\cite{tulving1972episodic}. It is characterized by: (i) \textbf{Egocentricity}. Episodic memory plays a crucial role in shaping our sense of self. Unlike semantic memory, which involves shared general knowledge, episodic memory is inherently self-referenced and unique to each individual~\cite{penaud2023role}. (ii) \textbf{Mental time travel}. Episodic memory helps us make decisions by allowing us to recall and relive moments from the past, guiding our choices in the present~\cite{tulving2002,nicholas2022uncertainty}. (iii) \textbf{Real-world convergence}. In the real world, experiences are continuous and countless, making it impossible to retain all details~\cite{neisser1992phantom}. However, episodic memory can always reliably store information about ``what'' happened, ``where'', and ``when''~\cite{yonelinas2015slow,Chandra2025}, thus condensing an infinite stream of observations into finite, discrete events. 
	Overall, in the context of AI, episodic memory offers specific advantages: (i) \textbf{Robustness}. The properties of attractors make EM models resistant to noise. (ii) \textbf{Interpretability}. The attractor state space corresponds to the event space, so a change in state represent shifts between events.
	%In essence, episodic memory not only enables the storage of an almost limitless array of subject-specific experiences but also facilitates their conscious retrieval to support decision-making.

	% EM 研究的优势， 可解释性+biologically plausibility
	% 鲁棒性
	
	%For example, no matter what specifics I may have noted, I distinctly remember that yersterday morning (when), I was in the office (where), engaging in a meeting (what). This unique form of memory, which allows us to consciously recall past experiences linked to particular places and times, is known as episodic memory \cite{tulving2002}.
	
	\begin{figure}[t]
		\centering
		\includegraphics[width=0.66\linewidth]{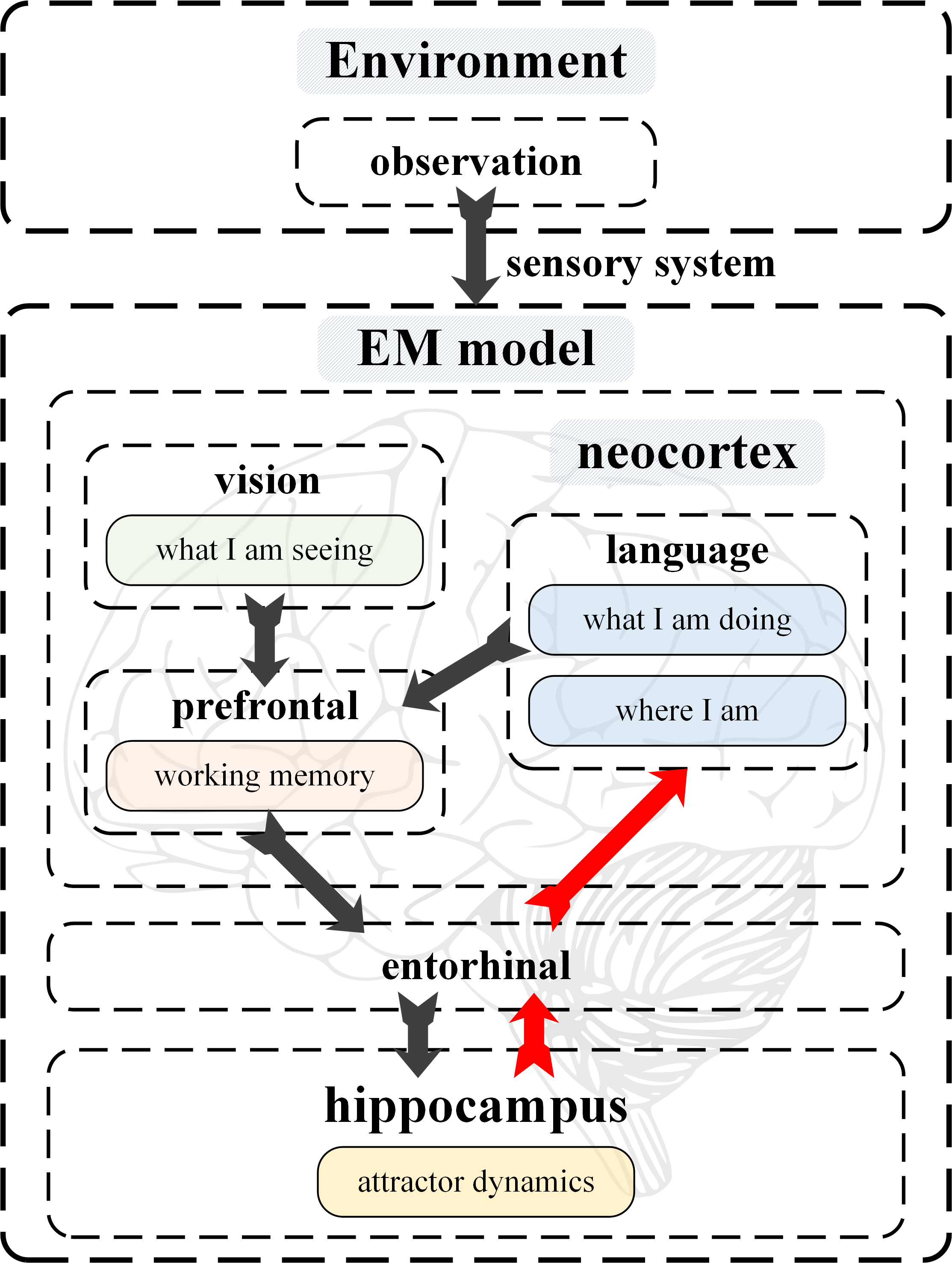}
		\vspace{-0.1in}
		\caption{\textbf{Diagram of Vision-Language Episodic Memory framework.}\label{fig:diagram} This diagram illustrates the biologically inspired structure of our episodic memory model. The ``vision'' and ``language'' components handle the perception of visual inputs and self-state descriptions. ``Working memory'' processes sequential inputs for short-term storage, while the ``entorhinal cortex'' acts as a gateway between the neocortex and hippocampus. The ``hippocampus'' manages episodic memory through attractor dynamics.}
		\vspace{-0.2in}
	\end{figure}
	
	Due to its unique characteristics, episodic memory has garnered significant attention across psychology, neuroscience, and artificial intelligence. Despite extensive research, the underlying mechanisms of episodic memory remain unclear, and there is no consensus on the optimal methods for modeling and effectively leveraging its capabilities. Originally proposed in psychology~\cite{tulving1972episodic}, early EM models were primarily designed to explain findings from psychological behavioral experiments~\cite{criss2015models}, such as the word frequency mirror effect~\cite{malmberg2004modeling}. These models focused on pattern matching and associations within tasks~\cite{humphreys1989different,gillund1984retrieval,anderson1977distinctive}. As neuroscience advanced, biologically inspired computational models became central to EM research, aiming to replicate its cognitive functions. Cognitive architectures, such as Soar~\cite{laird1987soar,laird2008extending}, incorporated episodic memory as a core component of overall cognitive processes, drawing insights from biological evidence~\cite{langley2009cognitive}. Further studies focused on the neural circuits involved in episodic memory, trying to build biologically realistic models that support its function~\cite{rolls2024theory}. Additionally, the key biological attractor dynamics in the CA3 region of the hippocampus have become widely accepted in neuroscience~\cite{rolls2024theory,allen2013evolution,squire1993structure,jeong2015episodic,rolls2018storage}. Attractor networks, such as Hopfield network~\cite{hopfield1982neural,hopfield2023}, have been used to mimic the convergent properties and neuronal dynamics of episodic memory. 
	%These advancements provide a solid foundation for modeling biologically plausible episodic memory in AI frameworks, which could potentially bring us closer to understanding its mechanisms and, for the first time, applying biological EM models in real-world situations.
	Thus, a deeper understanding of episodic memory through interpretable computational models can provide a solid foundation for integrating biologically plausible mechanisms into AI frameworks. This approach could bring us closer to understanding how these mechanisms work and, for the first time, enable the application of biological episodic memory models in real-world scenarios.
	
	%The interpretable mechanisms present a solid foundation for modeling biologically plausible episodic memory in AI frameworks, potentially advancing our understanding of its mechanisms and enabling the application of biological EM models in real-world situations for the first time.
	%These mechanisms provide a solid foundation for modeling biologically plausible episodic memory in AI frameworks, potentially advancing our grasp of its interpretability and enabling the application of biological EM models in real-world scenarios for the first time.

	% Thus, the deeper understanding of episodic memory through interpretable computational models provide a solid foundation for fusing biologically plausible into AI frameworks, 
	% the deeper understanding of mechanism of episodic memory 
	% provide a solid foundation for modeling biologically plausible episodic memory in AI frameworks, which could potentially bring us closer to understanding its mechanisms and, for the first time, applying biological EM models in real-world situations.

	% + advacements provides us great framework ... 提一下我们要做的事
	
	%However, limited efforts have been made to realize of EM in real-world simulations , which are crucial for practical applications.
	%
	%This paper focus on the real-world simulation for EM model. 
	
	Episodic memory is also considered crucial for AI, as it supports numerous critical high-level cognitive functions~\cite{hassabis2007deconstructing,eugilmez2015role}. Without the ability to remember past experiences, AI agents risk repeating previous mistakes and wasting valuable cognitive resources~\cite{jockel2008towards}. Additionally, the capacity to retrieve specific experiences is vital for fast learning in new or sparse-reward situations~\cite{allen2013evolution,boyle2024elements}. As a result, various AI research initiatives aim to glean insights from episodic memory to enrich AI agents
	%and bring them closer to replicating human-like intelligence
	~\cite{eugilmez2015role,jockel2008towards}. The Ego4d~\cite{EGO4D} project released a large-scale egocentric video dataset and EM benchmark, treating episodic memory as a special video modality, followed by studies that frame memory retrieval as a video question-answering task~\cite{Datta_2022_CVPR,Barmann_2022_CVPR}. 
	However, ``episodic memory'' here primarily describes systems that possess some features of it but differ in significant ways~\cite{boyle2024elements}.
	Therefore, it is crucial to introduce biological episodic memory mechanisms into AI models to improve their interpretability and robustness.

	%However, in this context, ``episodic memory'' refers only to systems incorporating certain aspects
	%lacking underlying mechanisms
	%of it
	%yet there are significant differences in the underlying mechanisms
	%~\cite{boyle2024elements}. Therefore, it is crucial to introduce biological episodic memory mechanisms into AI models to improve their interpretability and robustness.
	
	% 通过引入EM的神经科学模型 -> 可解释性+鲁棒性
	
	Clearly, there is a notable difference between EM models in neuroscience and AI. The former~\cite{hopfield1982neural,rolls2024theory} focuses on interpretable mechanism but faces challenges in practical application, while the latter~\cite{Datta_2022_CVPR,Barmann_2022_CVPR} prioritizes applicability but lacks interpretablity and robustness.
	
	%(+summarize... deeper into the mechanism to solve practical problem)

	%biological plausibility. 
	
	%Constructing a biologically plausible EM model for agents in realistic environments has been a challenging task for a long time. On the one hand, existing EM models in AI focus on limited 
	
	%In this paper, we aim at constructing a biologically plausible EM model for agents in realistic environments. On one hand, most existing EM models in AI focus on limited features of EM~\cite{boyle2024elements,Datta_2022_CVPR,Barmann_2022_CVPR}, often overlooking the critical biological attractor dynamics found in the CA3 region of the hippocampus, which play a key role in EM~\cite{rolls2024theory,allen2013evolution,squire1993structure,jeong2015episodic,rolls2018storage}. On the other hand, many studies~\cite{emmodel2024rolls,pmlr-v202-karuvally23a} restrict their evaluation to pattern-based data, which significantly limits their real-world applicability.

	%To address the challenaging task of episodic memory, 
	
	%{\Huge \textbf{\textit{!!!framework NAME!!!}}}
	
	%In our framework, 'where,' 'what,' and 'when' are modeled independently through distinct attractor networks, each representing a key aspect of episodic memory.
	
	%To Bridge this gap, we propose a novel decomposed episodic memory model 
	
	% -------- EpiGibson 放到后面去
	% overview -> vision+language -> hippocampus -> working memory & entorhinal
	
	To bridge this gap, we propose a novel Vision-Language Episodic Memory (VLEM) framework, which augment hippocampal attractor dynamics with large-scale pretrained model to create a biologically plausible episodic memory system within AI framework. Specifically, as illustrated in Fig.~\ref{fig:diagram}, we integrate the powerful semantic understanding capability of large-scale pretrained models~\cite{schuhmann2022laionb} to mimic the semantic processing in the cortex, transforming sensory input into semantic embeddings. With an understanding of the current state, the hippocampus supports episodic memory through its attractor mechanism, enabling stable storage and retrieval of experiences. Additionally, our framework incorporates working memory to track short-term historical states, while the entorhinal cortex collects information from the cortex and projects it back after interacting with the hippocampus, acting as a gateway between the two. The EM model maps an unlimited number of observations to a finite set of stable states based on its attractor properties, with gradient descent optimization used to learn the attractor space through end-to-end training. To ensure the EM model performs effectively in real-world scenarios, we have further developed EpiGibson, the first high-fidelity EM synthesis platform within a 3D physical simulation, built on OmniGibson~\cite{omnigibson}. Through our VLEM framework, we explore the construction of an EM model for a human-like agent operating in a physically realistic environment.
	
	%~\cite{schuhmann2022laionb}
	
	%propose a novel episodic memory framework that integrates an EM simulation environment, \textbf{EpiGibson}, with a biologically plausible EM model (as shown in Fig~\ref{fig:diagram}).  Additionally, 
	
	%we design a biologically plausible structure for the EM model, where ``vision'' and ``language'' represent the perception of visual input and self-state descriptions, ``working memory'' processes sequential perceptions for short-term memory, ``entorhinal'' serves as a gate between the neocortex and hippocampus, and the 
	% -------- hippocampus
	%``hippocampus'' is responsible for episodic memory~\cite{rolls2021brain}. Specifically, we utilize CLIP model~\cite{schuhmann2022laionb} for semantic processing in the cortex to extract vision and linguistic perceptions, RNN slots~\cite{workingmemory2024} to model working memory, and attractor RNNs~\cite{hopfield1984} to model episodic memory. The EM model maps an unlimited number of observations into a finite number of stable states based on its attractor properties, with gradient descent optimization used to learn the attractor space by end-to-end training. Through our framework, we explore the construction of the EM model for a human-like agent operating in a physically realistic environment.
	
	%In our experiments, + \textbf{we define the episodic memory task as ..., its significance ...}
	
	% + Task 的意义
	
	In our experiments, we tested the model on both pattern-based and simulation-based datasets. We reported its prediction performance during simulation, its robustness under noisy conditions, and its interpretability based on memory retrieval. Finally, we visualized the data in the simulation-based dataset to demonstrate the feasibility of applying our model in real-world scenarios. These results show that our VLEM framework can efficiently and reliably learn high-level temporal representations from an agent's sensory input in the environment.
	
	In summary, we have these contributions: 1) We introduce the VLEM framework, which combines large-scale pretrained models with hippocampal attractor dynamics, leveraging AI’s strong semantic understanding and biologically plausible episodic memory. 2) We present EpiGibson, a 3D physical simulation platform for generating episodic memory data, capable of simulating daily life and recording memory-related data. 3) We validate the robustness, interpretability, and real-world applicability of our framework through carefully designed experiments.

	\section{2.\ Methods}
	
	\begin{figure*}[t]
		\centering
		\includegraphics[width=0.7\linewidth]{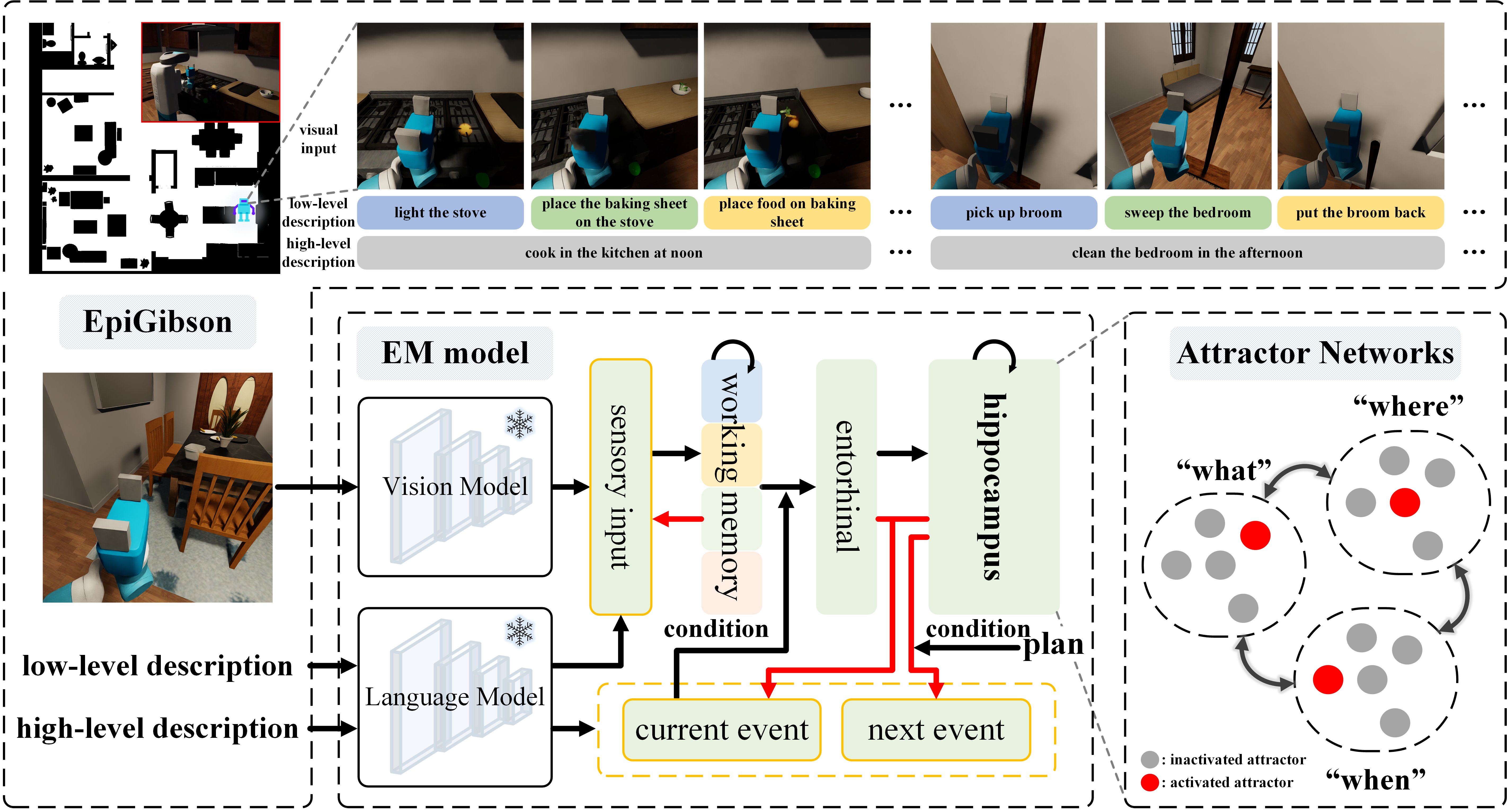}
		\vspace{-0.1in}
		\caption{\textbf{Details in VLEM framework.}\label{fig:method} \textbf{1) EpiGibson}: A simulation platform designed to evaluate the real-world applicability of our episodic memory model. It records continuous visual inputs and textual descriptions (``where'', ``what'', ``when'', and action) by sequentially performing events and actions in a 3D virtual environment, creating realistic datasets for model evaluation. \textbf{2) EM Model}:
%Visual input and text descriptions from EpiGibson are fed into vision and language models to generate sensory input and event embeddings. These sensory inputs are then projected into the working memory, where the slot closest to the current input is selected to encode current sensory input. The slots in working memory are unordered and equal, with their readout conditioned by the current event. This process creates an embedding of the short-term historical state, which is sent to the entorhinal cortex. As the gateway between the neocortex and hippocampus, the entorhinal cortex collects information from the neocortex and projects it back to maintain the current event or predict the next event after interacting with the hippocampus. The red arrows show backward projection to reproduce the last-layer embedding, while the yellow outline indicates the valid target for loss calculation. 
Vision and language models process visual and text inputs from EpiGibson to generate sensory and event embeddings, which are stored in working memory. The most relevant slot is chosen to encode the current sensory input. This short-term state is then passed to the entorhinal cortex, which connects the neocortex and hippocampus, helping to maintain or predict events. Red arrows show the backward projection, and the yellow outline marks the target for loss calculation.
\textbf{3) Attractor Networks}: The hippocampus is modeled as three interconnected attractor networks, representing the state of ``where'', ``what'' and ``when'' respectively, with event transitions represented by attractor state transitions.}
	\vspace{-0.15in}
	\end{figure*}
	
	%\subsection{Problem setup}
	
	In daily life, humans receive sensory inputs from the environment, which the brain processes to form an understanding of the self-state. Here, we categorize the self-state into low-level action descriptions (e.g., putting vegetables in a pan) and high-level event descriptions (e.g., cooking in the kitchen at noon). More specifically, we present the Vision-Language Episodic Memory framework, which combines hippocampal attractor dynamics with large-scale pretrained models to create a biologically inspired episodic memory system within AI framework. This framework processes sensory input into semantic embeddings, incorporates working memory and the entorhinal cortex for short-term memory and information gateway respectively, and uses the hippocampus to store and retrieve experiences.
%	 VLEM aims to build an episodic memory system for human-like agents in realistic environments.

	%we present the \textbf{VLEM} framework, which integrates the \textbf{EpiGibson} simulation environment with a biologically plausible EM model. \textbf{EpiGibson} serves as a high-fidelity platform for EM synthesis within a 3D simulation. The EM model incorporates components for vision and language perception, working memory (prefrontal cortex), entorhinal cortex, and hippocampal episodic memory~\cite{rolls2021brain}. We utilize CLIP model for semantic processing, RNN slots for working memory, and attractor RNNs for episodic memory. The framework’s attractor properties allow \textbf{EH-EM} to converge an unlimited number of observations to stable attractor states, with end-to-end optimization facilitating learning of attractor space.

	%Episodic memory contains attributions of what" happened, "where," and "when"~\cite{yonelinas2015slow}. Hence, we set two-level description for recoding 
	%
	%
	%We apply two method to synthesis data for evaluation.
	%
	%
	%Llama3 70B
	%
	%omniverse
	
	\subsection{2.1\ Data Synthesis}
	
	To better evaluate the robustness, interpretability, and real-world applicability of VLEM, we use two types of datasets: a pattern-based dataset, created with randomly generated patterns, and a simulation-based dataset, which records data from simulating events and actions with continuous visual input in our developed simulation environment, EpiGibson.
	
	\subsubsection{Dataset definition}
	
	The episodic memory model continuously generates high-level event descriptions from visual input and low-level action descriptions. As shown in Fig~\ref{fig:method}, the episodic memory dataset is defined as a combination of continuous visual input and low-level/high-level descriptions. Specifically, the dataset consists of a sequence of $L$ events $[E_1, E_2, \cdots, E_L]$, where each event $E_i$ corresponds to a description of what happened, denoted as $\mathrm{text}_{what,i}$. Each event $E_i$ also contains an action sequence $\mathbf{A}_i=[A_{i,1}, A_{i,2}, ..., A_{i,na_i}]$, representing the execution process of event $i$. Each action $A_{i,j}$ is linked to a low-level description $\mathrm{text}_{action,i,j}$, and includes a sensory sequence $\mathbf{S}_{i,j}=[S_{i,j,1}, S_{i,j,2}, ..., S_{i,j,ns_{ij}}]$, along with the corresponding ``where'' descriptions $\{\mathrm{text}_{where,i,j,k}\mid 1\leq k\leq ns_{ij}\}$ and ``when'' descriptions $\{\mathrm{text}_{when,i,j,k}\mid 1\leq k\leq ns_{ij}\}$. Overall, for each data sample $D_{i,j,k}$, sensory input is $S_{i,j,k}$, low-level description is $\mathrm{text}_{action,i,j}$ and high-level description is combination of $\mathrm{text}_{what,i}$, $\mathrm{text}_{where,i,j,k}$ and $\mathrm{text}_{when,i,j,k}$. Besides, there is a plan description $\mathrm{text}_{plan}$ summarizing the event sequence. These data samples are then flattened across the time dimension, with the data sample at time step $t$ represented as: $\hat D_t=\{S_t,\mathrm{text}_{action,t}, \mathrm{text}_{where,t}, \mathrm{text}_{what,t}, \mathrm{text}_{when,t}\}=D_{i,j,k}$. The $P_{plan}$ and $P_{when,t}$ are generated as random patterns that vary across different days and times of day, respectively. The other embeddings are then derived using pre-trained vision and language models:
	\begin{align}
		P_{sensory,t}&=\mathrm{Vision Model}(S_t)\label{eq1}\\
%		P_{plan}&=\mathrm{Language Model}(\mathrm{text}_{plan})\\
		P_{X,t}&=\mathrm{Language Model}(\mathrm{text}_{X,t}),\nonumber\\ &X\in\{where,what,action\}\label{eq3}
	\end{align}
	
	\subsection{2.2Vision-Languange Episodic Memory Framework}
	
	As shown in Fig.~\ref{fig:method}, our proposed EM framework consists of four modules: vision and language models, working memory, entorhinal cortex, and episodic memory. These modules represent key cognitive functions in the human brain. The modeling methods for each module are detailed below.
	
	\subsubsection{Vision and Language Models}
	
	With the rapid development of AI, large-scale pre-trained models now achieve human-level semantic understanding in areas like vision and language by learning from vast amounts of data.  While these models differ from the brain's biological mechanisms, research shows they effectively map data to semantic space~\cite{wang2022MMPTMSurvey}.  As a result, we use pre-trained vision and language models to simulate how the brain encodes the semantic understanding of vision and language, aiding the learning of working memory and episodic memory.
	
	Specifically, in our framework, we use the CLIP model~\cite{schuhmann2022laionb,Radford2021LearningTV,ilharco_gabriel_2021_5143773}, which aligns images and text in a shared semantic space using contrastive learning. This model encodes the brain's semantic understanding of visual inputs and self-state descriptions, mirroring how the neocortex processes this information. As shown in Eq. \ref{eq1}-\ref{eq3}, the vision and language models map image and text data into semantic space.
	
	\subsubsection{Working Memory}
	
	%	Working memory (WM), supported by the prefrontal cortex (PFC)~\cite{FUNAHASHIS1993Pnai}, enables the brain to temporarily store and manipulate information for short-term tasks. 
	%	Unlike episodic memory, which involves the hippocampus, WM relies on sustained neural activity instead of synaptic changes. 
	Previous studies suggest that WM can be represented as controllable activity slots, and RNN slots have been used to model WM with gradient descent optimization\cite{workingmemory2024}. Building on this idea, we further improved the model by adding loss functions to better align with WM mechanisms and a cross-attention-based readout process.
%	 This enabled us to successfully apply the working memory model in realistic simulations.
	
	We use $N_{slots}$ unordered RNNs to model each slot, with $N_{slots}=7$. Although some studies suggest a working memory capacity of 4 items~\cite{Cowan_2001}, we follow the more classical view of the ``magical number seven''~\cite{workingmemory7}. These slots are connected to each other through fully connected layers. Let the states of working memory slots as $\mathrm{WM}_i\in\mathbb{R}^{N_{WM}}, 1\leq i\leq N_{slots}$, where $N_{WM}$ is dimension of each working memory slot. The working memory iteration is then defined as: $\mathrm{WM}_{i,t+1}=\mathrm{tanh}(W_{input,i}P_{input,t}+\sum_{j=1}^{N_{slots}}W_{i,j}\mathrm{WM}_{j,t}+b_i)$. Here, $P_{input,t}=\mathrm{concat}(P_{sensory,t},P_{action,t})\in\mathbb{R}^{N_S+N_A}$ represents the sensory input dimension to working memory, while $W_{input,i},W_{i,j}$ are learnable weight matrices and $b_i$ is the bias term.
	
	Since all the RNN slots are identical in structure, it’s important to prevent them from storing identical embeddings. We define the similarity between two slots as: $\mathrm{sim}_{WM}(i,j)=\frac{1}{\parallel WM_i\parallel_2 \parallel WM_j\parallel_2}\mathrm{WM}_i\cdot\mathrm{WM}_j^T$,
	where $\parallel\cdot\parallel_2$ denotes L2 norm. 
	Because not all slots are activated at all times, we introduce an factor to represent the activation level of each slot, defined as: $a(i)=\parallel WM_i\parallel_2/\sqrt{N_{WM}}$. To encourage diversity between slots while allowing some slots to remain inactive, we derive the working memory loss function as: $\mathcal{L}_{WM}=\frac{1}{N_{slots}(N_{slots}-1)}\sum_{i\neq j}a(i)a(j)\big\vert\mathrm{sim}_{WM}(i,j)\big\vert\nonumber=\frac{1}{N_{WM}N_{slots}(N_{slots}-1)}\sum_{i\neq j} \big\vert\mathrm{WM}_i\cdot \mathrm{WM}_j^T\big\vert\label{eq:wmloss}$.

	\vspace{0.1in}
	\subsubsection{Entorhinal Cortex}
	The entorhinal cortex, acting as the gateway between the neocortex and hippocampus, collects information from working memory. While there are other pathways from specific brain regions to the entorhinal cortex, we simplify the model by ignoring these connections, as we assume that all necessary information can be encoded directly in working memory, theoretically from a modeling perspective. Therefore, the entorhinal state can be considered a readout of working memory, conditioned on the predicted current event embedding $\hat P_{curEvent,t}\in\mathbb{R}^{3N_P}$. The readout $\mathrm{Ento}_t\in\mathbb R^{N_{ento}}$ is calculated as:
		$\mathrm{Ento}_t=\mathrm{CrossAttn}(\hat P_{curEvent,t},\textbf{WM}_t)$.
	Here, $\mathrm{CrossAttn}(X_q,X_{kv})=\mathrm{softmax}(c\cdot Q\cdot K^T)\cdot V$, $Q=X_q\cdot W_q,\ K=X_{kv}\cdot W_k,\ V=X_{kv}\cdot W_v$, and $c=N_{ento}^{-0.5}$ is a constant scale factor.
	
	\subsubsection{Attractor Networks}
	
	%	The hippocampus plays a crucial role in episodic memory and exhibits attractor dynamics in its CA3 region~\cite{squire1993structure}. Hopfield network~\cite{hopfield1982neural}, a attractor network, are often used to model the convergent properties and neuronal dynamics associated with episodic memory. 
	In this study, we focus on modeling the CA3 region of the hippocampus. Based on the ability of episodic memory to stably recall the three attributes of an event\textemdash``where'', ``what'', and ``when''\textemdash we model each attribute with a separate attractor network and connect them to form a event attractor network. This allows us to build an attractor network that explicitly captures all the attributes of an event, offering a comprehensive model of the hippocampus.
	
	Let the attractor states of ``where'', ``what'' and ``when'' be denoted as $\mathrm{EM}_{where},\mathrm{EM}_{what},\mathrm{EM}_{when}\in\mathbb R^{N_{EM}}$, where $N_{EM}$ is the dimension of each attractor. Then attractor state of event is concatenation of them: $\mathrm{EM}_{event}=\mathrm{concat}(\mathrm{EM}_{where},\mathrm{EM}_{what},\mathrm{EM}_{when})\in\mathbb R^{3N_{EM}}$. Following the principles of the Hopfield network~\cite{hopfield1984}, we treat an RNN with symmetric recurrent weights as an attractor network. Let $\mathrm{attrs}=\{where,what,when\}$, the iteration for episodic memory is then given by:
	\begin{align*}
		\mathrm{EM}_{X,t,0}&=\tanh(\sum_{X'\in\mathrm{attrs}}W_{X,X'}\mathrm{EM}_{X',t}+W_{Ento}\mathrm{Ento}_t)\\
		\mathrm{EM}_{X,t,k+1}&=\tanh(\sum_{X'\in\mathrm{attrs}}W_{X,X'}\mathrm{EM}_{X',t,k})\\
		\mathrm{EM}_{X,t+1}&=\mathrm{EM}_{X,t,K},\ \ X\in \mathrm{attrs}
	\end{align*}
	where $W_{X,X'}$ is a symmetric matrix, $W_{X,X'}=W_{X',X}$, and $K$ is the number of additional self-iteration steps for the attractor network.

	Unlike previous approaches that rely on Hebbian learning, our framework uses gradient descent for model training, which means there is no need to predefine the attractor states. Instead, they can be learned directly from the data. In this case, we assume that the target attractor states can be predicted directly from the event states as: $\hat P_{EM,X,t}=\tanh(W_{event,X}P_{event,t}), X\in\mathrm{attrs}\label{event2em}$.
	In order to automatically learn these states as attractors, we define the attractor loss function as: $\mathcal{L}_{EM1}=\sum_{X\in\mathrm{attrs}}\parallel \hat P_{EM,X}-\tanh(W_{X,X}\hat{P}_{EM,X})\parallel_1\nonumber+\parallel \hat P_{EM,event}-\tanh(W_{event,event}\hat{P}_{EM,event})\parallel_1\label{lossem1}$,
	where $\hat P_{EM,event}$ represents the combined state of the three attractor states, and $W_{event,event}$ represents the recurrent connection weights after combining the three RNNs into one. In addition, to prevent all attractors from converging to the same state, we introduce an episodic memory contrastive loss as: $\mathcal{L}_{EM2}=\frac{1}{N_{event}(N_{event}-1)}\sum_{i\neq j}\mathrm{sim}_{EM}(i,j)\label{lossem2}$, 
	where $N_{event}$ denote the number of unique events and $\mathrm{sim}_{EM}(i,j)=\hat P_{EM,event,i}\cdot\hat P_{EM,event,j}^T / (\parallel \hat P_{EM,event,i}\parallel_2 \cdot \parallel \hat P_{EM,event,j}\parallel_2)$. Further, in order to ensure that neurons in attractor states are either fully firing or not firing at all, meaning their activation values are 1 or -1, thus enhancing the capacity of the network, we also introduce the following loss constraint: $\mathcal{L}_{EM3}=-\parallel\hat P_{EM,event}\parallel_1\label{lossem3}$.
	
	\subsubsection{Backward Projection}
	%	\subsubsection{Working Memory to Sensory Input} 
	1) From working memory to sensory input. To ensure that semantic information is encoded in working memory, we project the state of working memory back onto sensory input for prediction. Since working memory has multiple slots, with each slot encoding different semantic information, we designate the nearest slot as the one encoding the semantic information of the current sensory input. Therefore, the loss of encoding accuracy in working memory is as: $\mathcal{L}_{input}=\parallel \tanh(W_{input\_WM}P_{input})-\mathrm{WM}_k\parallel_2^2+\parallel W_{WM\_input}\mathrm{WM}_k-P_{input}\parallel_2^2\label{lossinput}$, where $k=\underset{k}{\mathrm{arg\,min}}\parallel \tanh(W_{input\_WM}P_{input})-\mathrm{WM}_k\parallel_2$.
	2) From hippocampus to events. To understand the current state of an event and predict future events, we project from the hippocampus back to the encoding of the event. Since the attractor states in the hippocampus do not directly encode semantic information, we predict the current understanding of the event by using the condition entorhinal state from the hippocampus state, and we predict the next event using the condition plan embedding. Thus we have: $\hat P_{event}=W_{EM\_event}\mathrm{EM}_{event} + W_{ento\_event}\mathrm{Ento}$ and $\hat P_{nxtEvent}=W_{EM\_nxtEvent}\mathrm{EM}_{event} + W_{plan\_event}P_{plan}$.
	Then prediction loss of events is derived as: $\mathcal{L}_{event}=\parallel P_{event}-\hat P_{event}\parallel_2^2+\parallel P_{nxtEvent}-\hat P_{nxtEvent}\parallel_2^2\label{lossevent}$.
	
	\subsection{2.3\ Training Strategy}
	
	We perform end-to-end training to optimize the EM model. The pretrained vision and language models are frozen, while all other weights are learnable. The loss used to optimize the entire model is combination of predefined losses:
	\begin{align*}
		\mathcal{L}=\mathcal L_{WM}+\mathcal L_{input}+\mathcal{L}_{event}+\alpha(\mathcal L_{EM1}+\mathcal L_{EM2}+\mathcal L_{EM3})
	\end{align*}
	where $\alpha$ is the scale of the EM-related loss. In order to enhance the learning performance of episodic memory, $\alpha$ will gradually increase as the training progresses.
	
	\section{3.\ Experiments}
	
	\input{tabs/results.tex}

	To more thoroughly assess the VLEM model's performance, we tested three attractor models in our experiments: (1) the Hopfield Network~\cite{hopfield1984}, (2) VLEM(merged), a version of VLEM with a single combined attractor, and (3) VLEM, the full VLEM model. In VLEM(merged), the three separate attractors for ``where'', ``what'', and ``when'' are combined into one, removing the explicit distinction between these categories. The Hopfield Network further extends VLEM(merged) by replacing the learned attractor weights with new ones through Hebbian learning.
	
	\subsection{3.1\ Datasets and Metrics}
	
	\subsubsection{Pattern-based Synthesis}
	
	In pattern-based synthesis, we construct a random tree graph where each node represents a unique location (``where''). A virtual agent completes its action list by navigating through the tree to execute actions at corresponding locations.	The agent begins at the location of the first action. Once the action is completed, it moves along a path to the next location to perform the following action. Transitioning to the next location and completing an action takes random time. For each time step, patterns for ``where'', ``what'' and ``when'' are recorded as $P_{where}$, $P_{what}$ and $P_{when}$, respectively, where $P_{where}, P_{what}, P_{when} \in \mathbb{R}^{T \times N_P}$, with $T$ being the total number of time points and $N_P$ being the dimension of each pattern. The embeddings for sensory input $P_{sensory,t} \in \mathbb{R}^{N_S}$ and low-level action description $P_{action,t} \in \mathbb{R}^{N_{A}}$ are calculated using two randomly initialized fully connected layers by passing $P_{event,t}$. The plan embedding is the average of all event embeddings: $P_{plan}=\mathrm{mean}(P_{event})\in \mathbb{R}^{3N_P}$.
	
	Let $N_{what}, N_{where}, N_{when}, N_{action}$ represent the number of unique patterns for ``what'', ``where'', ``when'' and actions, respectively. We create datasets with various settings for ($N_{what}, N_{where}, N_{when}, N_{action}$): ``large'' (50,20,10,100), ``medium'' (20,10,5,50) and ``small'' (10,5,3,20).
	
	\subsubsection{Simulation-based Synthesis}
	
	To further assess the real-world applicability of our model, we developed \textbf{EpiGibson}, the first episodic memory physical simulation platform, based on OmniGibson~\cite{omnigibson}. Like the pattern-based dataset, the simulation-based dataset is created by having the agent perform actions sequentially in each event, with each action's code manually programmed. During these interactions, the robot's visual inputs, along with the corresponding low-level and high-level textual descriptions, are recorded at each time step. As a result, the data format from the simulation-based dataset is the same as that of the pattern-based dataset. The key difference is that in the simulation-based synthesis, the agent continuously interacts with a 3D virtual environment within a physical simulation, providing a high-fidelity reproduction of human daily life while capturing the required data.
	
	Specifically, as shown in Fig. \ref{fig:method}, at each time step $t$, the data sample includes visual input $S_t\in\mathbb R^{H\times W\times 3}$, and text descriptions for ``where'', ``what'', ``when'' and action, denoted as $\mathrm{text}_{where,t}$, $\mathrm{text}_{what,t}$, $\mathrm{text}_{when,t}$ and $\mathrm{text}_{action,t}$ respectively. Furthermore, the event description is defined as a combination of ``where'', ``what'' and ``when'' descriptions. The action descriptions are then summarized to produce the plan description $\mathrm{text}_{plan}$ by ChatGPT-4o. Finally, all patterns are derived using the equations in Eq. \ref{eq1}-\ref{eq3}. 
	
	\subsubsection{Metrics}
	
	We used two metrics, MSE and correlation, to evaluate our predictions. We tested the predictions for the current event, the next event, and the sensory input separately.
	
	\subsubsection{Implementation Details}
	
	The learning rate starts at 2e-4, decaying every 500 steps, with training limited to 5,000 steps. All training and inference were done on a NVIDIA A800 GPU. The code is available at: \url{https://github.com/fudan-birlab/VLEM}.
	
	%	The learning rate starts at 2e-4 and decreases every 500 steps. Training is limited to 5,000 steps. All training and inference were done on a single NVIDIA A800 GPU. The code will be released after the paper is accepted.
	
	\subsection{3.2\ In-simulation Accuracy}
	
	To evaluate how well our model understands the state during agent simulation, we first assess its ability to predict sensory input, checking if the working memory has correctly stored the current sensory information. Further, we also evaluate the model’s predictions of the current and next events to see if the episodic memory retains the event details based on the input stimuli. Specifically, at each time step, the model receives an input $P_{sensory,t}$, and we test whether its predictions $\hat P_{sensory,t}$, $\hat P_{event,t}$ and $\hat P_{nxtEvent,t}$ are accurate. The results are shown in Tab.~\ref{tab:results}. VLEM significantly outperforms the Hopfield Network and surpasses VLEM(merged) on most metrics.
	
	\begin{figure}[t]
		\centering
		\includegraphics[width=0.9\linewidth]{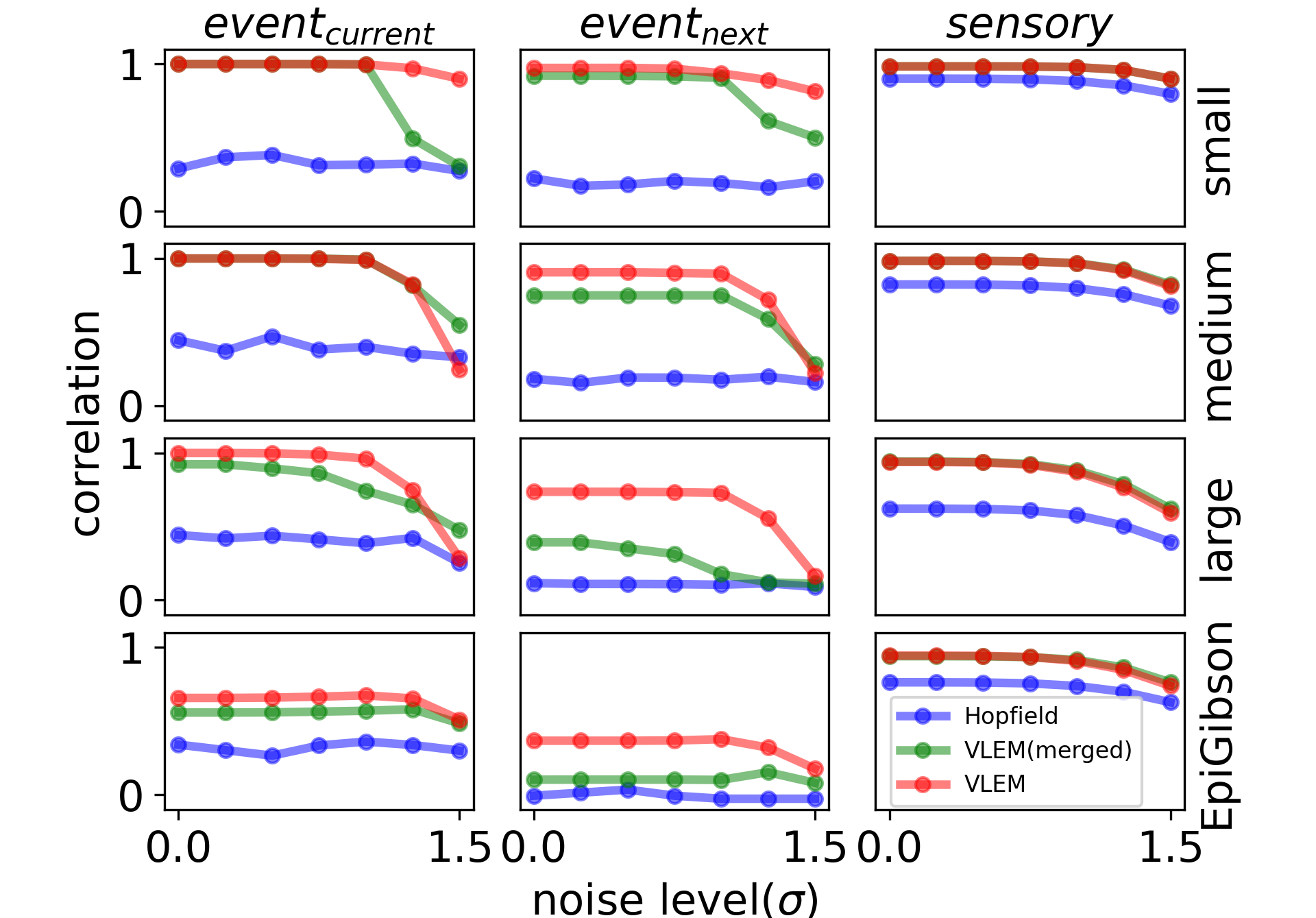}
		\vspace{-0.15in}
		\caption{\textbf{Evaluation on various datasets and noise levels.}\label{fig:eval-results} }
		\vspace{-0.15in}
	\end{figure}
	
	\subsection{3.3\ Robustness}
	
	To test the model's robustness, we evaluate its performance under varying levels of dataset and noise. As shown in Fig.~\ref{fig:eval-results}, for event prediction, as the number of events increases (from ``small'' to ``large''), VLEM's advantage over competitors becomes more evident, especially in predicting the next event. This suggests that explicitly splitting event elements (``where'', ``what'' and ``when'') and forming separate attractor networks increases the overall memory capacity, making the system more robust to larger data scales and capable of adapting to more complex environments. Additionally, we tested the model's robustness to input noise by adding varying levels of Gaussian noise ($\sigma: 0 \rightarrow 1.5$). The results show that VLEM is highly robust to input noise. Even with $\sigma=1$, VLEM maintains small accuracy losses (see tab.~\ref{tab:results}). The stability in sensory prediction accuracy highlights the robustness of the working memory model, with only limited impact from changes in episodic memory modeling. Overall, the experimental results demonstrate that VLEM is more robust to complex events and noisy inputs. 
	
	%\subsection{}
	
	%Since our model is based on the attractor properties of the hippocampus, we can retrieve a complete event using partial cues. For example, for an event with $P_{where}, P_{what}, P_{when}$, we can use the "where" and "when" information to retrieve the "what" information. More specifically, we obtain the attractor states $P_{EM,where}$, $P_{EM,what}$, and $P_{EM,when}$ through Eq.~\ref{event2em}, then fix two of these as state to the attractor network. We iterate the network without new input and compare the retrieved information for the remaining attribute.
	
	%The retuls is shown in ...

	\begin{figure}[t]
		\centering
		\includegraphics[width=\linewidth]{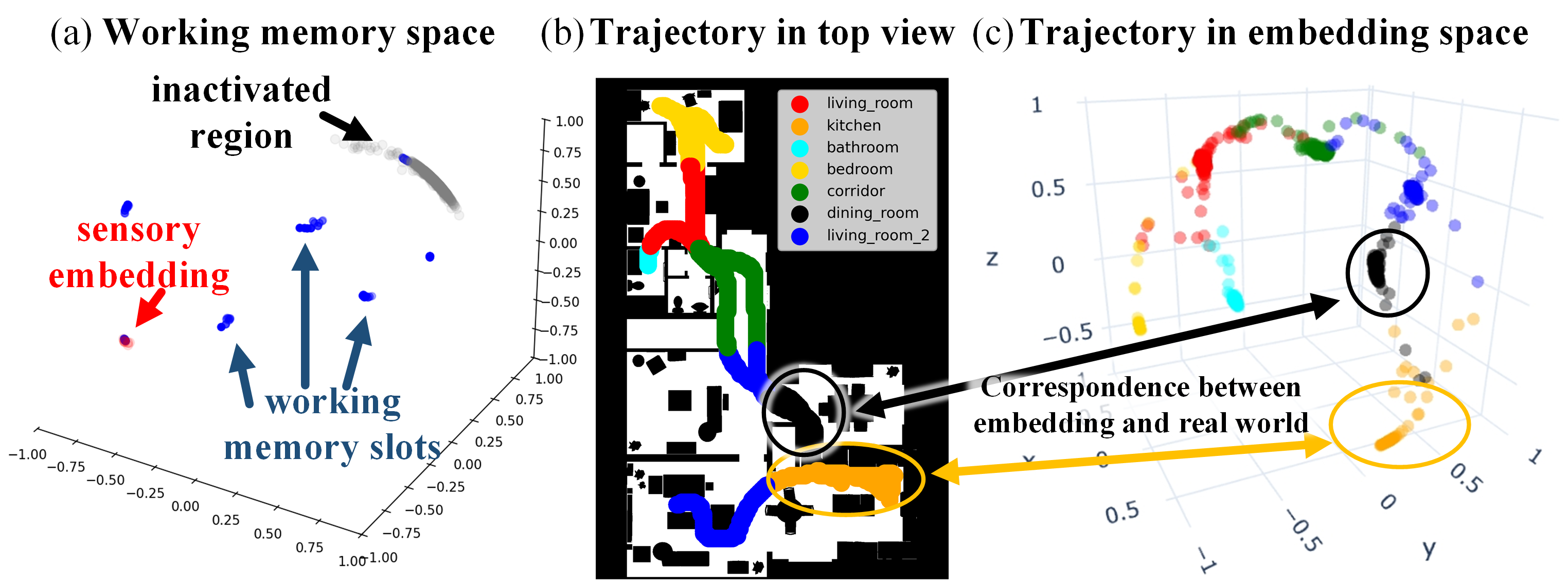}
		\vspace{-0.3in}
		\caption{\textbf{Visualization for working memory and episodic memory.}\label{fig:results} (a) Working memory slots either encode distinct semantics or remain inactive, with one specific slot effectively capturing the sensory input. (b) Real trajectory of agent in simulation environment. (c) The event (``where'') embeddings, derived from episodic memory, shows the agent's trace, with different colors representing different locations. This structure closely matches the real-world map.}
		\vspace{-0.2in}
	\end{figure}
	
	\subsection{3.4\ Real-world Applicability and Interpretability}
	
	To test our model's performance in real-world applications, we collected episodic memory data from a physical simulation using EpiGibson. In this setup, our visual input consists of real images, and the self-state descriptions are text labels corresponding to the agent’s state. We report the model's basic metrics on this dataset and visualize the results, comparing them with the actual physical labels. Results are shown in Tab.~\ref{tab:results}. Moreover, we use CEBRA~\cite{schneider2023cebra} for unsupervised encoding of neuronal dynamics into 3D space, as shown in Fig.~\ref{fig:results}. The results show that VLEM accurately learns spatial relationships consistent with the real world, effectively demonstrating the model's interpretability.
	
	\section{4.\ Conclusion}
	
	In this paper, we propose the Vision-Language Episodic Memory (VLEM) framework, which combines large-scale pretrained models with hippocampal attractor dynamics. The framework leverages AI's semantic understanding alongside the stability and interpretability of hippocampal dynamics, enabling reliable storage and retrieval of episodic experiences. We also present EpiGibson, a 3D simulation platform for generating episodic memory data, and show how the framework is robust, interpretable, and applicable in real-world scenarios. Our work advances biologically inspired memory models and their integration into AI systems.

	%	\nocite{ChalnickBillman1988a}
	%	\nocite{Feigenbaum1963a}
	%	\nocite{Hill1983a}
	%	\nocite{OhlssonLangley1985a}
	%	% \nocite{Lewis1978a}
	%	\nocite{Matlock2001}
	%	\nocite{NewellSimon1972a}
	%	\nocite{ShragerLangley1990a}

	\bibliographystyle{apacite}
	
	\setlength{\bibleftmargin}{.125in}
	\setlength{\bibindent}{-\bibleftmargin}
	
	\bibliography{CogSci_Template}

\end{document}

%% file: tabs/results.tex
% \usepackage{multirow}
\setlength{\tabcolsep}{5pt}
\begin{table*}[]
	\centering
\begin{tabular}{cccccccccccccc}
	\hline
	\multicolumn{1}{l}{}      & \multicolumn{1}{l}{} & \multicolumn{6}{c}{pattern-based dataset (large)}                                          & \multicolumn{6}{c}{simulation-based dataset}                                                      \\ \cline{3-14} 
	&                      & \multicolumn{2}{c}{VLEM}        & \multicolumn{2}{c}{VLEM(merged)} & \multicolumn{2}{c}{Hopfield} & \multicolumn{2}{c}{VLEM}        & \multicolumn{2}{c}{VLEM(merged)} & \multicolumn{2}{c}{Hopfield} \\ \cline{3-14} 
	& $\sigma$             & corr           & MSE            & corr            & MSE            & corr          & MSE          & corr           & MSE            & corr            & MSE            & corr          & MSE          \\ \hline
	\multirow{2}{*}{curEvent} & 0                    & \textbf{0.999} & \textbf{0.001} & 0.922           & 0.136          & 0.443         & 0.851        & \textbf{0.657} & \textbf{0.670} & 0.558           & 0.804          & 0.366         & 1.044        \\
	& 1                    & \textbf{0.961} & \textbf{0.077} & 0.743           & 0.440          & 0.387         & 0.897        & \textbf{0.674} & \textbf{0.618} & 0.570           & 0.746          & 0.377         & 1.030        \\ \hline
	\multirow{2}{*}{nxtEvent} & 0                    & \textbf{0.736} & \textbf{0.450} & 0.393           & 0.926          & 0.116         & 1.194        & \textbf{0.366} & \textbf{1.044} & 0.102           & 1.179          & -0.007        & 1.341        \\
	& 1                    & \textbf{0.729} & \textbf{0.461} & 0.177           & 1.314          & 0.105         & 1.267        & \textbf{0.377} & \textbf{1.030} & 0.100           & 1.181          & -0.028        & 1.365        \\ \hline
	\multirow{2}{*}{sensory}  & 0                    & 0.939          & 0.117          & \textbf{0.941}  & \textbf{0.113} & 0.622         & 0.649        & \textbf{0.944} & \textbf{0.108} & 0.940           & 0.115          & 0.763         & 0.418        \\
	& 1                    & 0.871          & 0.240          & \textbf{0.882}  & \textbf{0.221} & 0.577         & 0.718        & 0.909          & 0.174          & \textbf{0.915}  & \textbf{0.162} & 0.738         & 0.456        \\ \hline
\end{tabular}
	\vspace{-0.1in}
	\caption{\textbf{Evaluation results for accuracy across synthetic datasets.}\label{tab:results} We evaluated our models on both pattern-based and simulation-based datasets, using metrics to assess predictions for the current event, next event, and sensory input. Our VLEM significantly outperforms Hopfield and demonstrates robustness even under Gaussian noise with a standard deviation of $\sigma=1$. Bold values indicate the best results.}
	\vspace{-0.2in}
\end{table*}